\documentclass{article}
\usepackage{authblk}
\usepackage{cite}
\usepackage{amsmath,amssymb,amsfonts}
\usepackage{soul}
\usepackage{algorithmic}
\usepackage{graphicx}
\usepackage{textcomp}
\usepackage{xcolor}
\def\BibTeX{{\rm B\kern-.05em{\sc i\kern-.025em b}\kern-.08em
    T\kern-.1667em\lower.7ex\hbox{E}\kern-.125emX}}
\begin{document}

\title{Modular multi--source prediction of drug side--effects with DruGNN}

\author[1,2,C]{Pietro Bongini}
\author[2]{Franco Scarselli}
\author[2]{Monica Bianchini}
\author[2]{Giovanna Maria Dimitri}
\author[1,2]{Niccol{\`o} Pancino}
\author[3]{Pietro Li{\`o}}
\affil[1]{Department of Information Engineering, University of Florence, Florence, Italy}
\affil[2]{Department of Information Engineering and Mathematics, University of Siena, Siena, Italy}
\affil[3]{Department of Computer Science and Technology, University of Cambridge, Cambridge, United Kingdom}
\affil[C]{Corresponding author: pietro.bongini@gmail.com}

\maketitle

\begin{abstract}
Drug Side--Effects (DSEs) have a high impact on public health, care system costs, and drug discovery processes. Predicting the probability of side--effects, before their occurrence, is fundamental to reduce this impact, in particular on drug discovery. Candidate molecules could be screened before undergoing clinical trials, reducing the costs in time, money, and health of the participants. Drug side--effects are triggered by complex biological processes involving many different entities, from drug structures to protein--protein interactions. To predict their occurrence, it is necessary to integrate data from heterogeneous sources. In this work, such heterogeneous data is integrated into a graph dataset, expressively representing the relational information between different entities, such as drug molecules and genes. The relational nature of the dataset represents an important novelty for drug side--effect predictors. Graph Neural Networks (GNNs) are exploited to predict DSEs on our dataset with very promising results. GNNs are deep learning models that can process graph--structured data, with minimal information loss, and have been applied on a wide variety of biological tasks. Our experimental results confirm the advantage of using relationships between data entities, suggesting interesting future developments in this scope. The experimentation also shows the importance of specific subsets of data in determining associations between drugs and side--effects.
\end{abstract}

\section*{Keywords}
\noindent
Graph Neural Networks, Drugs side--effect prediction, Node classification, Deep learning, Transductive learning.

\section{Introduction} \label{section_introduction}
Drug Side--Effects (DSEs) represent a common health risk, with an estimated 3.5\% of all hospital admissions, and approximately 197,000 annual deaths, in Europe alone, related to adverse drug reactions \cite{ADR_Numbers}. Such adverse outcomes turn out to be extremely expensive for public care systems. Drug--related morbidity and mortality are estimated to have cost nearly $177.4$ billion in the United States alone in the year 2000 \cite{ADR_Costs}. As prescription drug use is increasing \cite{ADR_Increase}, the numbers and costs related to DSEs are also expected to rise. DSEs are a huge problem for pharmaceutical companies, as their occurrence during clinical trials slows down drug discovery processes and prevents many candidate molecules from being selected as commercial drugs \cite{ADR_drug_discovery}. Therefore, predicting DSEs before submitting a molecule to clinical trials is extremely important to avoid health risks for participants and cut drug development costs \cite{Predictor_Mizutani}. 
\\
Computational prediction methods, and in particular deep learning methods, are techniques of growing importance in this scope \cite{Predictor_Deepside}. DSEs are, in fact, triggered by complex biological mechanisms, involving interactions between different entities, such as drug functional groups, proteins, genes, and metabolic processes. As a consequence, an efficient predictor should be capable of processing heterogeneous data, accounting for the relationships among different data types \cite{Polypharmacy_Deac}. In the last decade, DSE computational prediction methods have evolved from simple predictors based on euclidean data \cite{Predictor_Mizutani} or drug similarity \cite{Predictor_Zhang} to Machine Learning (ML) methods based on Support Vector Machines \cite{Predictor_Metabolic} or clustering \cite{Predictor_Drugclust}, and to more complex predictors based on Random Forests \cite{Predictor_RF} or deep learning \cite{Predictor_Deepside}. Current machine learning methods for DSE prediction have increased the number and variety of features considered for computing the predictions, but they are still widely based on euclidean (vectorial) data, whereas the relevant information for DSE prediction is relational in nature. This is a limit, since the relational information must undergo a  preprocessing to be transformed into vectors, with an inevitable loss of information. In addition, preprocessing methods usually require re--thinking when new features are added.
\\
In recent years, Graph Neural Networks (GNNs) \cite{GNN_model} have become a solid standard for predicting and generating graph--structured data, thanks to their capability of processing relational data directly in graph form, with minimal loss of information, and in a flexible way \cite{GNN_review_properties}. After their introduction in 2005 \cite{GNN_theory,GNN_capabilities}, many different models have been added to the GNN family, e.g.,  Graph Convolution Networks (GCNs) \cite{GCN_standard}, spectral GCNs \cite{GCN_spectral_Bruna} \cite{GCN_spectral_Defferrard}, GraphSAGE \cite{GraphSAGE}, GraphNets \cite{GraphNets}, Message--Passing Neural Networks \cite{MPNN}, and Graph Attention Networks \cite{GAT}, just to mention the most important ones. 
\\
Models of the GNN family have repeatedly proved to be more efficient and accurate than non--graph--based predictors on many node, edge, and graph property prediction tasks. Moreover, GNNs have been employed in a wide variety of biological and chemical tasks \cite{GNN_review_applications}, also including several drug--discovery related problems \cite{GNN_review_drug_discovery}. In particular, the GNN model used in the present study has been employed for the prediction of protein--protein interactions \cite{GNN_implementation}, and for the generation of molecular graphs of potential new drug candidates \cite{MG2N2}.
\\
In this work, we propose a new method for single--drug side--effect prediction based on GNNs, and we build a graph dataset for this task, accounting for drug--gene, drug--drug, and gene--gene relationships. To the best of our knowledge, this is the first machine learning approach to be able  to exploit directly graph structured relational data, for the prediction of single--drug side--effects. GNNs were already used for a related but different task, namely the prediction of polypharmacy side--effects, both by analyzing the network of drugs and protein targets \cite{Polypharmacy_Decagon}, and by applying a graph co--attention model over the two graphs describing a pair of drugs' structural formulas \cite{Polypharmacy_Deac}. Nevertheless, as far as we know, no single--drug side--effect predictor based on GNNs has been proposed yet. Moreover, we believe that predicting a single drug side--effect could answer an immediate research question, namely, which are the expected side--effects of this new candidate drug?
\\
The main contributions of the paper are as follows.
\begin{itemize}
\item The first contribution of this work consists in the construction of a relational dataset for the prediction of DSEs, made with data coming from well--known publicly accessible resources. The dataset is a single heterogeneous graph, in which two types of nodes (drugs and genes) share three types of edges (drug--gene, drug--drug, and gene--gene relationships). Both drug and gene nodes have features, accounting respectively for their chemical properties and for their characteristics and function.
\item The second contribution of our work consists in a GNN--based method, called DruGNN, for the prediction of DSEs on the new dataset we constructed. The prediction is set up as a multi--class multi--label node classification problem (applied only to drug nodes, and not to gene nodes), in which each DSE corresponds to a class. We adopt a mixed inductive--transductive learning scheme \cite{GNN_transduction_theory}, that exploits both the features of drugs and genes (induction path) and the information on the side--effects of known drugs (transduction path), in order to predict the side--effects of new drugs. The whole method is flexible, since the graph dataset can be easily extended to include other node features and further relationships  without changing the machine learning framework \cite{GNN_transduction_application}.
\item 
The approach has been assessed experimentally, with very promising results, showing a good classification accuracy. The performance of DruGNN are compared to those of similar graph--based models (using the same inductive--transductive scheme) and to those of a deep Multi--Layer Perceptron (MLP) that cannot exploit relational information. Finally, two ablation studies, one over the set of side--effects, and the other over the set of features, show the model robustness and the contribution to learning brought by each single data source.
\item The usability of DruGNN is discussed, as it can be exploited for the prediction of DSEs of new drugs without retraining. It is sufficient to add a new compound, or a batch of new compounds, each represented by a new graph node, and to predict their classes exploiting the same inductive--transductive learning scheme exploited for training and testing. Interesting future developments in this direction are also analyzed: the same approach could be replicated on a tissue--specific basis, by exploiting tissue--specific transcriptomics and DSE information.
\end{itemize}
The rest of this paper is organized as follows: Section \ref{section_dataset} describes the dataset, its construction process, and the data sources; Section \ref{section_method} sketches the GNN--based prediction method; Section \ref{section_results} presents the results we obtained, and a discussion on their relevance and meaning; Section \ref{section_usability} discusses the expected use of our method; Section \ref{section_conclusions} draws conclusions on this work and summarizes the main results obtained. A more detailed description of the GNN model is provided in the Appendix.

\section{Dataset}
\label{section_dataset}

\begin{figure*}
\centering
\includegraphics[width=0.6\textwidth]{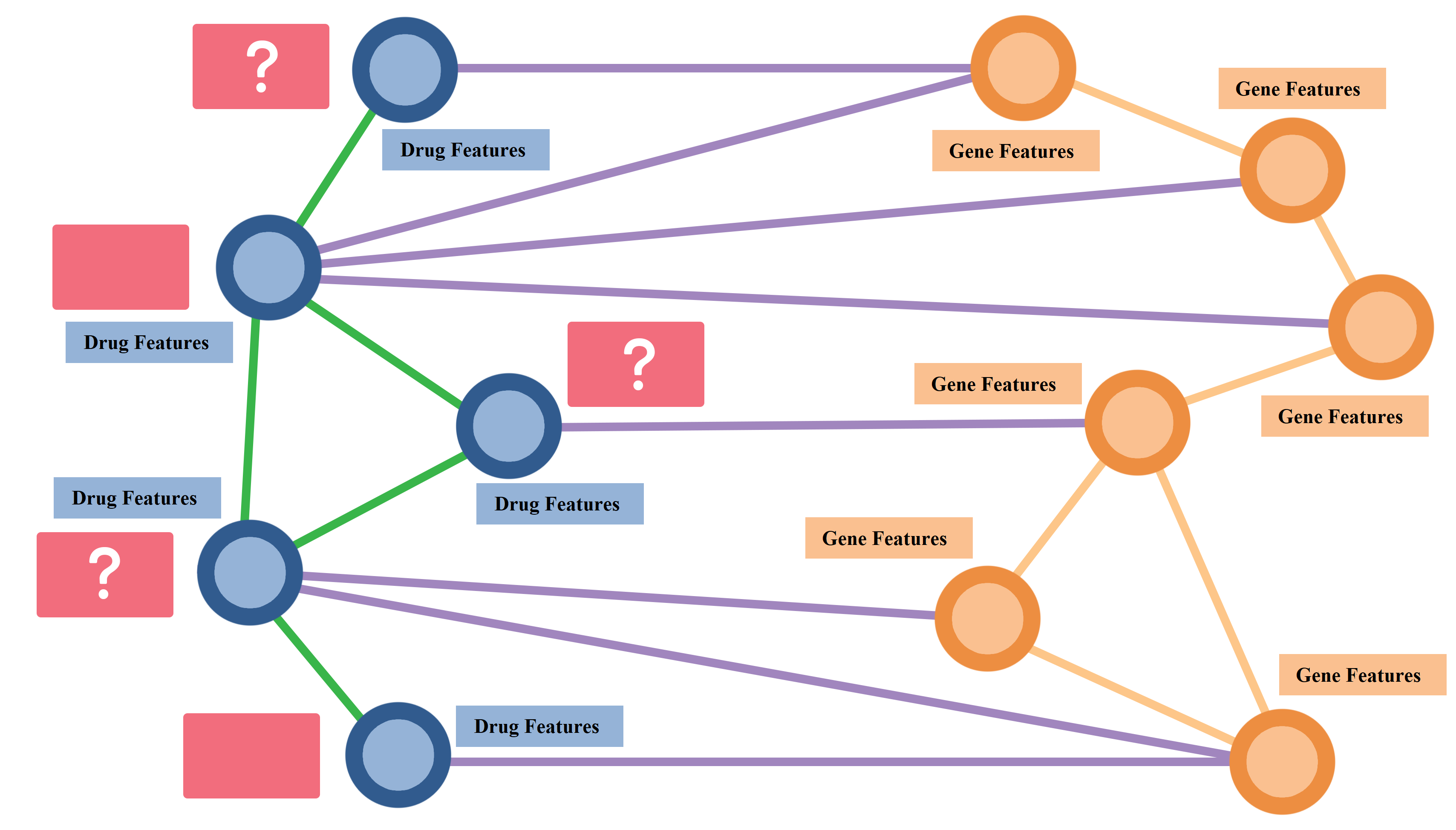}
\caption{Illustration of the graph composition. Drug nodes are represented as blue coloured circles, while gene nodes are represented as orange coloured circles. Red rectangles represent classes.}
\label{fig_graph}
\end{figure*}

\begin{figure*}
\includegraphics[width=\textwidth]{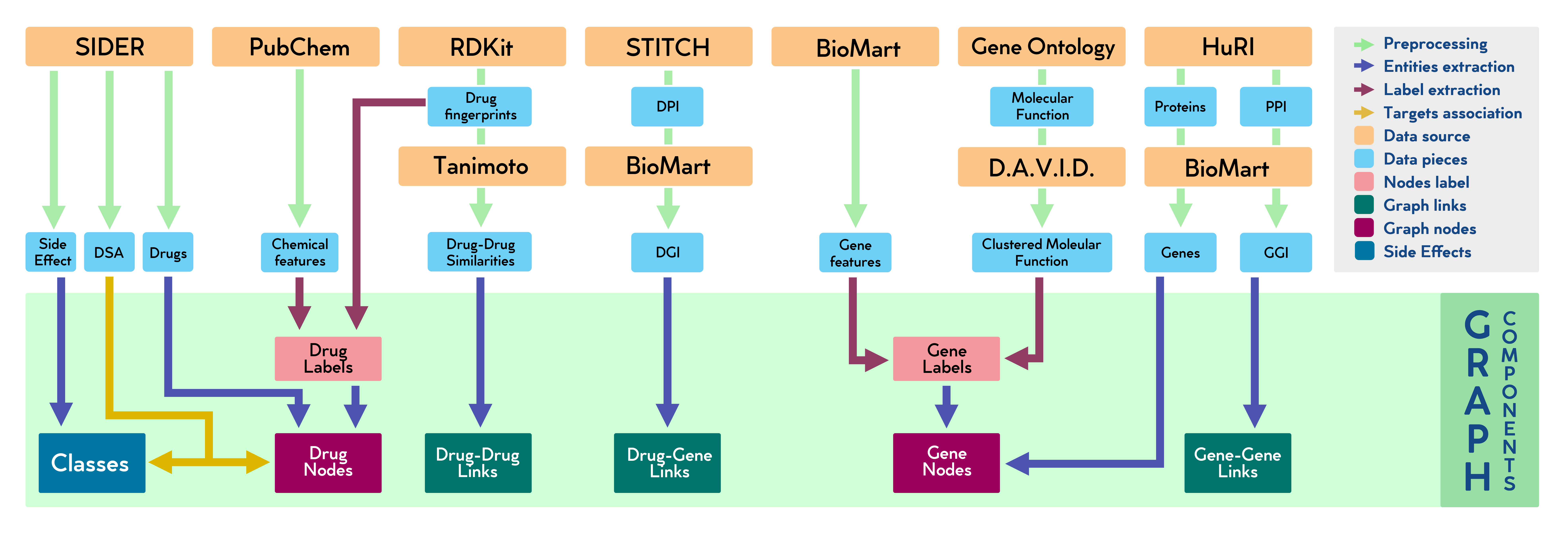}
\caption{Sketch of the dataset construction. Each data source is represented by an orange rectangle. Cyan rectangles represent data pieces. Preprocessing steps are represented by green arrows, which can include feeding data in input to other sources to obtain refined data. Graph node subsets are represented by purple rectangles, with their labels sketched as pink rectangles. Green rectangles are subsets of graph edges, while the blue rectangle represents the classes (side--effects). Red arrows represent the composition of feature labels from data pieces, while blue arrows show the composition of graph entities (nodes, edges, classes). The yellow arrow represents the association of drug nodes to side--effect classes.}
\label{fig_dataset}
\end{figure*}

Computational methods for the prediction of DSEs have mainly relied on euclidean derived features so far. Even methods, like  \cite{Predictor_Yamanishi}, that do use topological information (i.e. about the metabolic network), compress it into a euclidean space before processing. Since DSEs are triggered by complex biological phenomena, data for predicting DSEs are heterogeneous and come from multiple sources. Drug protein targets are of key importance, as highlighted by the good results of Sparse Canonical Correlation Analysis between drug targets and DSEs \cite{Predictor_Mizutani}. Chemical drug features play an important role too \cite{Predictor_Pauwels}, as well as metabolic data \cite{Predictor_Drugclust}. Combining all these pieces of information, even in euclidean form, yields the best results when using deep learning predictors \cite{Predictor_Deepside}. As a consequence, to build our dataset, we integrated information from all of these sources. The main novelty of our approach consists in building a graph with these data, and processing the graph as it is, without forcing data objects into euclidean vectors of features. 
\\ 
Our dataset consists of a single graph, in which each drug, as well as each gene, is mapped to a node. Both drug nodes and gene nodes are described by feature vectors. Edges represent drug--drug relationships, drug--gene interactions, and gene--gene interactions. Side--effect labels are associated to each drug node. These label will be used, according to the inductive--transductive scheme, as either transductive features for known drugs, or class supervisions for new drugs. A sketch of the graph is provided in Figure \ref{fig_graph}.
\\
The associations between drugs and side--effects were downloaded from the SIDER database \cite{Dataset_Sider}, which collects DSE information by aggregating multiple public information sources, summing up to 5,868 side--effects occurring on 1,430 drugs, with a total of 139,756 entries, each accounting for the association of a single drug to a specific side--effect. In our graph, a node was created for each drug. Each side--effect corresponds to a class. Our set of gene nodes, as well as the gene--gene edges, representing the interactions between two genes or their products, were constructed by downloading protein--protein interactions (PPI) information from the Human Reference Interactome (HuRI) \cite{Dataset_Huri}, and mapping each protein to the gene it is a product of. The product--gene associations were instead obtained from Biomart \cite{Tool_Biomart}. Drug--protein interactions (DPI) were downloaded from the STITCH database \cite{Dataset_Stitch}, one of the most complete and up--to--date DPI databases available. Once again, using Biomart, each protein was mapped to the gene it is a product of, obtaining the links between drug nodes and gene nodes.
\\
Drug features were retrieved from PubChem \cite{Dataset_Pubchem}, which provides seven chemical descriptors for each molecule in our dataset, as well as the SMILES string describing its structure. The seven chemical descriptors consist in: molecular weight (MW), polar surface area (PA), xlogp coefficient (LP), heavy atom count (AC), number of hydrogen bond donors (HD), number of hydrogen bond acceptors (HA), and number of rotatable bonds (RB). In order to better describe each drug molecule, we also translated its SMILES representation to the corresponding structural formula, and extracted its substructure fingerprint, using RDKit software\footnote{RDKit: Open-Source Cheminformatics Software, by Greg Landrum. URL: https://www.rdkit.org/}. In order to keep the feature vector size of drug nodes similar to that of gene nodes (gene feature vectors are 140--dimensional and will be described in the following), we opted for drug substructure fingerprints of size 128, bringing the total size of the drug feature vector to 135.
\\
Drug substructure fingerprints were also exploited to build the drug--drug set of edges, accounting for similarity relationships between molecules. In particular, we measured the Tanimoto similarity \cite{Tanimoto_Similarity} of the fingerprints of each pair of drugs, adding an edge only to those pairs which were above a similarity threshold (which was set as a hyperparameter at graph construction). Fingerprints were extracted with RDKit again, but with size 2048, in order to better estimate the Tanimoto similarity.
\\
Gene features were obtained from two sources. Biomart \cite{Tool_Biomart} provided 3 pieces of information: the chromosome, which was one--hot encoded for a total of 25 features (22 regular chromosomes, plus X, Y, and mitochondrial DNA); the strand the gene is codified on ($+1$ or $-1$); the percentage of GC content. Gene ontology \cite{Dataset_Gene_Ontology} provided the molecular function ontology terms to which each gene is mapped. These describe the molecular function of each gene, which, combined to the gene--gene interaction links, allow to reconstruct the metabolic network. Since Gene Ontology mapped our genes to a total of 3,422 terms, we clustered the terms to those appearing at the higher levels of the molecular function ontology, using DAVID \cite{Tool_David_A} \cite{Tool_David_B}. This produced 113 unique terms, which were one--hot encoded and concatenated to the gene features obtained from Biomart (for a total of 140 features on each gene node).
\\
We subsequently selected only side--effects with a sufficient number of occurrences in SIDER: In order for the network to be able to learn the associations of each side--effect, we applied a minimum threshold of 100 occurrences, reducing the number of side--effects in our dataset to 360. After this first filtering step, drugs without side--effects were also removed, reducing the number of drug nodes in the graph to 1,341. Genes with incomplete features were also discarded, along with their gene--gene interactions, bringing the number of gene nodes to 7,881. All the drugs have complete feature vectors, and at least one DPI. DPIs and DSEs of removed drugs were also removed. Drugs are mapped to 360 classes (one for each side--effect), with 96,477 total positive occurrences, and 515,160 negative ones. In this sense, belonging to the positive class for a drug means that it produces the particular side--effect. The target for each drug must be evaluated in relation to every possible side effect (360) since the addressed problem is both multi--class and multi--label. A total of 331,623 edges is instead present in the final version of the graph: 12,002 gene--gene interaction links, 314,369 DPI, and 5,252 drug--drug similarity links (with a minimum Tanimoto threshold of 0.7). The dataset construction, with all the source databases and preprocessing steps, is sketched in Figure \ref{fig_dataset}.

\section{Method}
\label{section_method}
Graph Neural Networks (GNNs) have seen an important and steady development in the last decade. Their main feature is the capability of processing graph--structured data with minimal information loss \cite{GNN_capabilities}. In our work, we exploit the original GNN model \cite{GNN_model}, and in particular its most recent implementation \cite{GNN_implementation}, to build a DSE predictor called DruGNN. An overview of the GNN model is provided in the Appendix of this paper. The final task is to predict the label of drug nodes only, solving a node--based classification problem, with multiple classes (360 side--effects) and in a multi--label setting (each drug can cause multiple side--effects). In order to provide repeatable and comparable results, we set a dataset split and always use that split throughout the experimentation. The test set contains 10\% of the drug nodes and is fed to the network only at test time. In our experiments, we also retain a 10\% of the drug nodes as a validation set, in order to check overfitting and stop the training procedure when this occurs. The rest of the nodes (80\%) is exploited as a training set.
\\
In our work, we make use of a mixed inductive--transductive learning scheme \cite{GNN_transduction_application}, in which the network learns the node--class associations exploiting a double mechanism. In standard inductive learning, the GNN model would predict the side--effects of drugs based on the node features of drugs and genes, and the graph connectivity. In a transductive learning setup, the GNN model would make the predictions based on the known side--effects of other drugs. In our mixed  inductive--transductive learning scheme, the GNN model exploits both mechanisms at the same time. 
\\
The learning scheme applied in this work consists in splitting the training set into ten batches. The network learns the input--supervision association on one training batch at a time, while the other nine batches are exploited as a transduction set. The features of each drug node in the transduction set are augmented with the transductive features, corresponding to the occurrence of the 360 side effects on that node. When analyzing the validation set, the full training set is exploited, in the same way as described before, as the transduction set. When analyzing the test set, the transduction set is composed of both the validation set and the training set.
\\
This scheme is particularly appropriate for the expected use of our dataset and tool: the idea is to exploit the known DSE associations to predict the DSEs of newly inserted drugs, and the mixed inductive--transductive scheme simulates this behaviour at training, validation, and test times.
\\
The network hyperparameters were tuned with an extensive grid--search over the validation set. In particular, we analyzed all the hyperparameter values described in Table \ref{table_hyperparameters} and their combinations. Each element in the grid was analyzed by measuring the average model accuracy in a training/validation experiment with five repetitions.

\begin{table}[!htb]
\begin{center}
\begin{tabular}{|c|c|c|}
\hline
\textbf{Hyperparameter} & \textbf{Values}& \textbf{Best Config.} 
\\
\hline
Activation&relu, selu, tanh, sigmoid&relu
\\
Initial Learning Rate&$10^{-2}$,$10^{-3}$,$10^{-4}$&$10^{-3}$ 
\\
State Dimension&$10$,$50$,$100$,$200$&$50$ 
\\
Hidden Units&$100$,$200$,$500$&$200$ 
\\
Neighborhood Aggregation&average, sum&average
\\
Dropout Rate&$0.5$, $0.3$,$0.1$,$0.0$&$0.0$
\\
\hline
\end{tabular}
\vspace*{0.2cm}
\caption{Hyperparameter values analyzed during the grid search procedure, and best configuration obtained.}
\label{table_hyperparameters}
\end{center}
\end{table}

After tuning the hyperparameters, in order to check the learning capabilities of the DruGNN on our dataset, and in particular the effect on the learning process of the reduction or expansion of the set of side--effects, we set up a dedicated series of experiments. In this part of the experimentation, which consists in an ablation study over the set of side--effects, our model was trained and tested on versions of our dataset with progressively reduced numbers of side--effects: only the most $k$ common side--effects were retained, with $k$ assuming values $\{360, 240, 120, 80, 40, 20, 10, 5\}$. 
\\
To evaluate the importance of the contributions of the different data sources, we carried out another ablation study. We grouped the features and the edges by source and eliminated one feature/edge group at a time from the dataset, evaluating the performance of the model in absence of that group. The performance gap obtained gives an estimate of the importance of the features that were kept out. There are seven feature/edge groups in our dataset, each of which was analyzed in an experiment repeated five times. Once again, we always used the same dataset split and the same transductive learning scheme described for the previous experimentation.
\\
Eventually, the DruGNN was compared to other competitive GNN models with different characteristics, in order to assess its performance with respect to the alternative solutions. In particular we focused on two powerful models: GCNs  \cite{GCN_standard}, which exploit convolutions to aggregate information coming from different locations across the graph --- and have shown competitive performance on many different tasks; GraphSAGE \cite{GraphSAGE}, which  are versatile networks that can be configured with various aggregation and state updating functions --- being potentially competitive on every graph dataset. Additionally, we also compared to a simple Multi--Layer Perceptron (MLP), in order to assess the difference between a graph--based model and a euclidean predictor. It was not possible to include previously published DSE predictors in the comparison, as our dataset is completely novel, and graph--structured, making it impossible to adapt to the feature sets of the predictors available in the literature (we remind that no graph--based predictor was published for this task). In particular, after a small optimization over the validation set, we used a three--layered MLP.

\section{Results and Discussion}
\label{section_results}
The hyperparameter search described in Section \ref{section_method} produced a model with an accuracy over the validation set of 87.22\%. The same model, evaluated on the held--out test set obtained an accuracy of 86.30\%.
\\
Given the best model configuration obtained in this first set of experiments, we investigated the contribution of the side--effects to the learning capability of the network. We ranked the side--effects by occurrences, and then we progressively reduced the size of the set of side--effects, by selecting only the most common ones. The average accuracy over five repetitions was measured over the held--out test set. Results are reported in Table \ref{table_DSE_ablation}.

\begin{table}[!htb]
\begin{center}
\begin{tabular}{|c|c|c|c|c|c|c|c|c|}
\hline
\textbf{DSE} & 360 & 240 & 120 & 80 & 40 & 20 & 10 & 5
\\
\hline
\textbf{Acc.\%} & $86.3$ & $81.5$ & $73.2$ & $68.5$ & $63.0$ & $61.8$ & $67.1$ & $74.7$
\\
\textbf{Bal.\%} & $58.1$ & $58.6$ & $60.0$ & $60.5$ & $62.1$ & $59.9$ & $57.7$ & $56.2$
\\
\hline
\end{tabular}
\vspace*{0.2cm}
\caption{Average accuracy percentage (Acc.\%), and average balanced accuracy percentage (Bal.\%) obtained on the test set by training and testing the model on progressively smaller sets of side--effects (DSEs).}
\label{table_DSE_ablation}
\end{center}
\end{table}

\noindent
Since we are dealing with a multi--class multi--label classification task, each class membership can be seen as a problem to be learned independently and in parallel with respect to all the other classes. As a consequence, the first expectation would be that increasing the number of classes, the network would have to learn a more complex algorithm, needing to solve more problems in parallel. On the contrary, the results reported in Table \ref{table_DSE_ablation} show a clear tendency of improvement of the performance for larger sets of side--effects. This counter--intuitive behaviour is due to the network ability of learning intermediate solutions, which are useful for all or large subsets of the classes, with an effect very similar to transfer learning. 
\\ 
This is particularly evident in our system, in which transfer learning between classes is fundamental because of the relatively small dimension of the set of drugs, with the additional bonus of avoiding overfitting. An inversion of this behaviour can be observed at lower set dimensions (up to 20), where transfer learning becomes less easy and convenient and the network learns to treat each class independently. 
\\
A second ablation study was carried out on the feature/edge groups coming from different data sources. The accuracy of the model, trained and tested in absence of the data group, was evaluated and averaged over five repetitions of the same experiment. Since the groups of features are of different sizes, to better weigh the importance of each, we also measured the DPF (Difference Per Feature) score: this is the performance difference with respect to the complete model, divided by the number of features in the group. The description, and the corresponding performance loss observed in the ablation study, of each data group, are reported in Table \ref{table_features}.

\begin{table}[!htb]
\begin{center}
\begin{tabular}{|c|c|c|c|c|c|c|c|}
\hline
\textbf{Group} & DPI & PPI & DDS & FP & PC & GO & BM
\\
\hline
\textbf{Type} & E & E & E & DF & DF & GF & GF
\\
\hline
\textbf{Acc.\%} & $86.14$ & $86.18$ & $86.23$ & $85.81$ & $86.20$ & $86.17$ & $86.20$
\\
\textbf{Diff.} & $0.16$ & $0.12$ & $0.07$ & $0.49$ & $0.10$ & $0.13$ & $0.10$
\\
\hline
\textbf{Count} & - & - & - & 128 & 7 & 113 & 27
\\
\textbf{DPF} & - & - & - & $0.004$ & $0.014$ & $0.001$ & $0.004$
\\
\hline
\end{tabular}
\vspace*{0.2cm}
\caption{Average accuracy percentage (Acc.\%), and difference with respect to the model trained on the complete feature set (Diff.), obtained on the test set by training and testing the model on our dataset in absence of the corresponding feature/edge group (Group). Each group is associated to a type: E (Edges), DF (Drug Features), GF (Gene Features). For DF and GF data groups, the number of features (Count) and the difference per feature (DPF) are reported: the latter is obtained by dividing the difference (Diff.) by the number of features in the group (Count). Our data groups are: DPI (Drug--Protein Interactions), PPI (Protein--Protein Interactions), DDS (Drug--Drug Similarity), FP (FingerPrints), PC (PubChem), GO (Gene Ontology), and BM (BioMart).}
\label{table_features}
\end{center}
\end{table}

Table \ref{table_features} shows that each data source has a positive contribution on the GNN learning process. In particular, deleting the drug fingerprints brings the largest performance drop, which can be explained by the importance of the drug substructures in determining the side--effects, but also by the large number of features (128) assigned to this data group.\\ Proportionally, looking at the DPF score, the seven PubChem descriptors have the highest contribution, as it could be expected given their chemical relevance. The gene features also have a relevant impact on performance, with the Biomart derived features having a DPF equal to that of drug fingerprints. Edges also showed to be important, as deleting each edge set leads to a performance drop. Instead results suggest that drug similarity relations are the less important, likely because drug similarity can be inferred by the network on the basis of the fingerprints and of the drug--gene interactions.
\\
Although each group of features and edges has a positive contribution to model performance, the small performance drop obtained by switching them off tells us that the model is robust. In fact, it works almost as well as the complete version even when entire sets of edges or features are deleted. We can therefore hypothesize the following. On the one hand, GNNs are expected to be robust, on the basis of previous systematic ablation studies that demonstrated their capabilities on many types of graph datasets \cite{GNN_power}. On the other hand, the large quantity of features and edges, and the heterogeneous nature of our data sources, likely boost the model's robustness. 
\\
Moreover, to assess the capabilities of DruGNN with respect to other GNN variants, and with respect to non--graph--based euclidean models, we carried out a comparison with GraphSage \cite{GraphSAGE}, GCNs \cite{GCN_standard}, and with a simple Multi--Layer Perceptron (MLP) model trained on a vectorized version of our drug data. The MLP gives a measure of the results that can be achieved by applying a traditional euclidean predictor on our dataset. The GCN and the GraphSage are trained with the same transductive scheme as DruGNN.  All the models were trained with the binary cross--entropy loss function, Adam optimizer \cite{Adam}, and an initial learning rate equal to $10^{-4}$. A maximum of $500$ epochs was allowed for each model, with early stopping on the validation loss, and recovery of the best weights. As expected, all the graph--based models outperformed the standard MLP, showing the advantage given by representing the dataset as relationships on a graph and by learning directly on the graph structure. Moreover, using a GraphSAGE or a GCN approach on this task did not allow to reach the same results we obtained with DruGNN, as shown in Table \ref{table_comparison}. This can be explained by the fact that our model is particularly efficient on node property prediction tasks, as the one presented in the current paper, while the other GNN models tend to aggregate nodes on a larger scale, getting an advantage on graph property prediction tasks. This is also in line with theoretical studies on GNNs that demonstrate the processing capabilities by simulating the Weisfeiler--Lehman test \cite{GNN_power}. Model evaluation is based on the average accuracy percentage obtained over 10 runs of training and testing on the same dataset split.

\begin{table}[!htb]
\begin{center}
\begin{tabular}{|c|c|c|}
\hline
\textbf{Model} & \textbf{Configuration}& \textbf{Avg. Acc. \%} 
\\
\hline
DruGNN & $K=6, SD=50, DL=1 \times 200$ & $86.30 \%$
\\
GCN & $CL=2 \times 36, DL=116$ & $82.94 \%$ 
\\
GraphSAGE & $CL=2 \times 72, DL=1 \times 168$ & $83.11 \%$ 
\\
MLP & $DL=3 \times 25$ & $77.98 \%$
\\
\hline
\end{tabular}
\vspace*{0.2cm}
\caption{Comparison between different models of the GNN family. Model configuration is reported, all of the models were optimized with a small hyperparameter search. K: maximum number of state update iterations for DruGNN; SD: state dimension for DruGNN; DL: number of dense layers and units in each dense layer; CL: number of convolutional layers and units in each convolutional layer. For the GCN and the GraphSage, the dense layer is the last one before the output layer.}
\label{table_comparison}
\end{center}
\end{table}

\section{Usability of DruGNN in real practice}
\label{section_usability}
DruGNN is meant as a tool of real usage, that can help healthcare and pharmacology professionals to predict side--effects of newly discovered drugs or other compounds not yet classified as commercial drugs. The dataset and the software are publicly available on github \footnote{https://github.com/PietroMSB/DrugSideEffects}, so that both assets can be exploited in further scientific research and by the whole community. Furthermore, both the dataset and the algorithm are scalable: adding new compounds to predict their side--effects does not compromise the network usability (i.e., the network does not need to be retrained from scratch).
\\
An example of such usage is represented by the prediction of the side--effects of Amoxicillin (PubChem CID: 2171), which is part of the held--out test set (and therefore never seen during the training or validation phases). Amoxicillin has been determined to be similar to the following drugs, listed by PubChem CID: 2173, 2349, 2559, 4607, 4730, 4834, 8982, 15232, 22502, 6437075. It also interacts with 76 genes. No other information but the fingerprint and PubChem features of Amoxicillin are available to the model. The network correctly predicts the following side--effects, listed by the SIDER id: C0000737 (Abdominal pain), C0001824 (Agranulocytosis), C0002792 (Anaphylactic shock), C0002871 (Anaemia), C0002878 (Haemolytic Anaemia), C0003467 (Anxiety), C0006840 (Candida infection), C0011991 (Diarrhoea), C0012833 (Dizziness), C0013378 (Dysgeusia), C0015230 (Rash), C0017178 (Gastrointestinal disorder), C0018681 (Headache), C0019080 (Haemorrhage), C0027497 (Nausea), C0033774 (Pruritus), C0038362 (Stomatitis), C0042075 (Urinary tract disorder), C0042109 (Urticaria), C0042963 (Vomiting), C0267792 (Hepatobiliary disease), C0917801 (Insomnia). It fails to predict these side--effects: C0002994 (Angioedema), C0008370 (Cholestasis), C0009319 (Colitis), C0011606 (Dermatitis exfoliative), C0014457 (Eosinophilia), C0036572 (Convulsion). Please notice that Angioedema, Colitis, Dermatitis exfoliative, and Convulsion are indicated as very rare for Amoxicillin. Cholestasis has relatively few occurrences in the dataset, and is therefore difficult to predict. Moreover, the network shows good predictive capabilities on side--effects which are common in the whole drug class Amoxicillin belongs to (represented by the similar compounds in the dataset). In addition, the network predicts only one side--effect which is not associated to Amoxicillin in the supervision: C0035078 (Renal failure).
\\
As shown in the example, to predict the side--effects of a new compound, it is sufficient to retrieve information (coming from wet--lab studies and from the literature) on its interactions with genes, and to know its structural formula. RdKit can be used to calculate the fingerprint, and consequently the similarity to other drugs in the dataset. The PubChem features can either be obtained from a database, or calculated with RdKit. It is then sufficient to insert the compound in the dataset and to predict its side--effects with DruGNN. Visualisation of the DruGNN results and the excepts from the database could then be directly used by doctors and pharmacists.

\section{Conclusions}
\label{section_conclusions}
Combining data from multiple sources is crucial for a deep neural network to learn complex mechanisms regulating the occurrence of drug side--effects. In particular, the relational information on the interactions of drugs and genes is well described by a graph structure. Integrating these entities and their relations, we built a graph dataset thought for training and testing graph--based DSE predictors. Graph Neural Networks (GNNs) showed very good learning capabilities on this dataset, suggesting that a predictor based on GNNs could help anticipate the occurrence of side--effects. Furthermore, its application on new candidate drugs would help saving time and money in drug discovery studies, also preventing health issues for the participants to the clinical tests.
\\
DruGNN is a modular approach to DSE prediction and is robust to ablation. Moreover, it is easily usable on new drug compounds: it is sufficient to add the new drug, with its features and gene interactions, as a node in the graph, and to run the prediction of its classes. The model does not need retraining, and the same inductive--transductive learning scheme can be used for future additions of compounds and predictions of their side--effects. The prediction relies on a modular multi--omics robust approach, based on information retrieved from publicly available sources. In principle, the same graph could be exploited also to predict the drug--gene interactions of new compounds, by applying link prediction over the gene set.
\\
Since drug structures proved to be one of the most important parts of our dataset for deep learning, an interesting future direction is represented by the development of a GNN--based predictor that could analyse the structural formulas of the molecules, represented as graphs. These molecular graphs could be augmented with features coming from the gene side and drug--gene relations. In this scope, the algorithm could even be combined with generative models, like MG\textsuperscript{2}N\textsuperscript{2} \cite{MG2N2}, that generate molecular graphs of possible drug candidates in large quantities. The task of the DSE predictor would be to screen out all the candidate compounds with high probabilities of occurrence of particular side--effects.
\\
Another very interesting direction is that of specializing the predictor presented in this work, in order to take into account tissue--specific data (i.e. gene expression) and fine--tune a dedicated version of the model for each tissue. This could be made possible by exploiting tissue specific side--effect targets, leading to a more detailed prediction which could also be personalized, given the gene expression values of each individual, as expected in the context of precision medicine.

\begin{appendix}
	\section*{Appendix: The Graph Neural Network Model}
	In this Appendix, we provide a brief description of the Graph Neural Network (GNN) model used in this work. For a more extensive overview, see the seminal papers \cite{GNN_theory}\cite{GNN_model}, and the work in which the current implementation was introduced \cite{GNN_implementation}.
	\\
	GNNs process graph--structured data, by replicating the graph topology in an encoding network. To build the encoding network, two Multi--Layer Perceptrons (MLPs) are used as building blocks and replicated over the graph. The state updating network is replicated over each node and implements the state transition function $f_{w}$, described in Eq. (\ref{eq_state}). The state is updated $K$ times, where $K$ is a network hyperparameter. At each iteration $t$, the state $x_{n}^{t}$ of a node $n$ is updated on the basis of its feature label $l_{n}$, its state at the previous iteration $x_{n}^{t-1}$, and the labels and states of all the nodes $m \in Ne(n)$, where $Ne(n)$ is the set of neighbours of $n$:
	\begin{equation}
	\label{eq_state}
	x_{n}^{t} = f_{w}(x_{n}^{t-1}, l_{n}, \ A(\{(x_{m}^{t-1}, l_{m}): m \in Ne(n)\})).
	\end{equation}
	
	The function $A$ can be any aggregation function of the neighbours of node $n$. In particular, we use the average of the node states and labels $A_{avg}$, and the sum of the node states and labels $A_{sum}$. A hyperparameter is used to select which aggregation to use, as described in Eq. (\ref{eq_aggregation}):
	\begin{align}
		\begin{aligned}
			\label{eq_aggregation}
			A_{avg} &= \frac{1}{|Ne(n)|}\sum_{m \in Ne(n)} (x_{m}^{t-1}, l_{m})
			\\
			A_{sum} &= \sum_{m \in Ne(n)} (x_{m}^{t-1}, l_{m}).
		\end{aligned}
	\end{align}
	
	Since the aim is to process a heterogeneous graph, in which we have two types of nodes (drugs and genes), we exploit two MLPs, one for each node type, for implementing $f_{w}$, as described in Eq. (\ref{eq_state}). During the training phase, these two MLPs will learn two different versions of $f_{w}$. Namely, $f_{w}^{d}$ will be applied to compute the states of all drug nodes, while $f_{w}^{g}$ will calculate the states of all gene nodes.
	\\
	After the $K$ iterations of state update are completed, the output network is applied to the subset of nodes in the graph for which an output is requested. In our case, these correspond to the drug nodes. The output network implements a function $g_{w}$, that calculates the output $\hat{y}_{n}$ for each output node $n$, taking its state $x_{n}^{K}$ and its feature label $l_{n}$ as inputs, as described in Eq. (\ref{eq_output}):
	
	\begin{equation}
	\label{eq_output}
	\hat{y}_{n} = g_{w}(x_{n}^{K}, l_{n}).
	\end{equation}
	
\begin{figure*}
		\includegraphics[width=\textwidth]{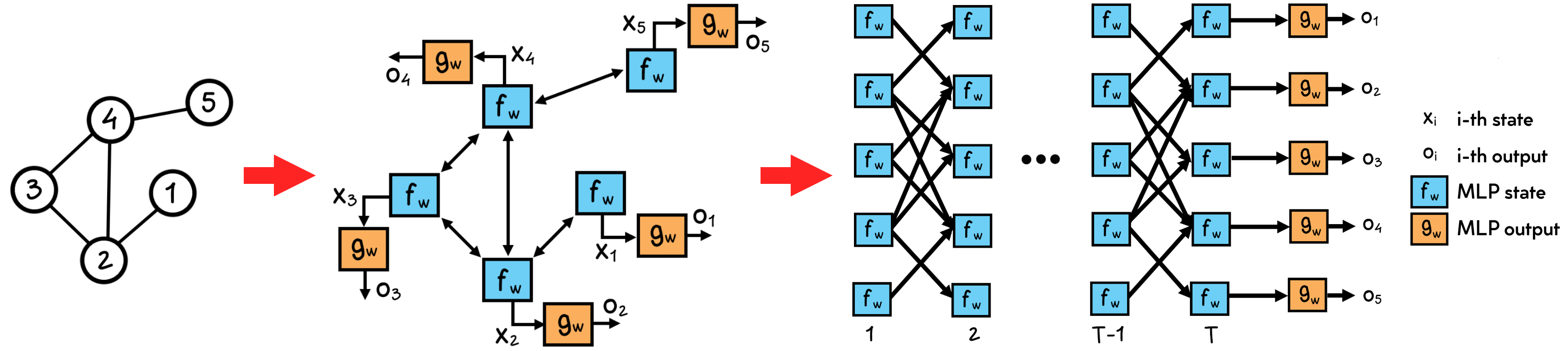}
		\caption{Sketch of how the GNN model produces the encoding network and its unfolded version on an example input graph.}
		\label{fig_network}
	\end{figure*}

	Unfolding the encoding network, taking into account the $K$ state iterations and the output calculation, we obtain a deep architecture, as illustrated in Figure \ref{fig_network}. Usually, the state updating network and the output network are MLPs with one hidden layer (as it is our case), but they can be deeper: with $L_{f}$ and $L_{g}$ layers respectively. As a consequence, the unfolded version of the network will have $L_{g} + L_{f} \times K$ layers. This means that, during backpropagation, the network layers closer to the input layer could suffer from the vanishing gradient phenomenon. As we stated before, though, all the repetitions in space (over each node of the graph) and time (over the $K$ iterations) of the state updating functions are replicas of the same $nt$ MLPs, where $nt$ is the number of types of nodes ($nt = 2$ in our case). This means that the GNN benefits from weight sharing throughout the graph topology and the iterations, allowing to learn complex functions over the deep architecture of the unfolded encoding network \cite{GNN_capabilities}.
	
\end{appendix}

\section*{Acknowledgments}
We declare no known conflict of interest concerning this work. This research did not receive specific funding.

\bibliographystyle{ieeetr}
\bibliography{references}

\end{document}